\documentclass[envcountsect,runningheads,a4paper]{llncs} 

\usepackage{amssymb,stmaryrd,amsmath}
\usepackage{mdwlist} 
\usepackage{float}   

\usepackage[colorlinks]{hyperref}

\spnewtheorem{thrm}{Theorem}[section]{\sc }{} 
\spnewtheorem{lemm}[thrm]{Lemma}{\sc }{} 
\spnewtheorem{prop}[thrm]{Proposition}{\sc }{} 
\spnewtheorem{corr}[thrm]{Corollary}{\sc }{} 

\spnewtheorem{nttn}[thrm]{Notation}{\sc }{}
\spnewtheorem{defn}[thrm]{Definition}{\sc }{}
\spnewtheorem{xmpl}[thrm]{Example}{\sc }{}
\spnewtheorem{rmrk}[thrm]{Remark}{\sc }{}

\usepackage{color}
\definecolor{mygreen}{rgb}{0,0.6,0}
\definecolor{mygray}{rgb}{0.5,0.5,0.5}
\definecolor{mymauve}{rgb}{0.58,0,0.82}

\usepackage{listings}
 
\definecolor{gray}{RGB}{128, 128, 128}
\definecolor{lightgray}{RGB}{200, 200, 200}
\definecolor{cyan}{RGB}{0, 255, 255}
\definecolor{blue}{RGB}{0, 0, 255}
\definecolor{red}{RGB}{255, 0, 0}
\definecolor{pink}{RGB}{255, 128, 128}
\definecolor{green}{RGB}{0, 128, 0}
\definecolor{lightyellow}{RGB}{255, 255, 200}
\definecolor{purple}{RGB}{128, 0, 128}

\newcommand\weblink[1]{\href{#1}{\emph{Weblink, accessed 9 July 2020}}}

\begin{document}
\title{Karl Marx and the Blockchain}
\author{Devraj Basu\inst{1},  Murdoch J. Gabbay\inst{2}}
\institute{Strathclyde University, Scotland, UK \and Heriot-Watt University, Scotland, UK}
\date{}
\maketitle
\begin{abstract}
Blockchain is often presented as a technological development; however, clearly it is not only that: the `Blockchain buzz' exists in the context of current social and political developments. 
In this essay, we analyse blockchain technology and its social and political context from a perspective of Marxist economic theory. 

Since arguably the last great inflection point in society and technology was analysed by Marx in terms of labour and capital and since we seem to be experiencing a shift in the balance between these
forces today, it makes sense to revisit the Marxist ideas and apply them to the current situation, to see how well they still apply and if necessary to update them for current events.
\end{abstract}
\thispagestyle{empty}

\section{Bitcoin and Distributed Ledger Technology}
\label{sect.bitcoin.and.dlt}

\subsection{Consider Blockchain} 

In 1991 blockchain was just a research idea \cite{haber:howtsd}, which was commercialised as a digital timestamping service called \emph{Surety} in 1995.\footnote{It's still running.}  
In 2008 a design for a cryptocurrency called Bitcoin was introduced by Satoshi Nakamoto (an alias); it went live in January 2009.  

On 10 October 2010 Bitcoin traded at 10 cents / bitcoin.  
By 29 February 2011 it was at 1 USD / Bitcoin.  By 1 April 2013 it reached 100 USD / Bitcoin.  
Bitcoin scraped 1000 USD / Bitcoin at the end of 2013, fell back, but then rallied and reached 19,783 USD / Bitcoin on 17 December 2017.  
Unsurprisingly there came a crash --- but Bitcoin was not wiped out and its reputation was not utterly destroyed: it survived and at time of writing, Bitcoin trades at around 10,000 USD / bitcoin (for a trading volume of roughly ten million transactions per month). 
That is a 100,000-fold increase in value in ten years.

It's not just as a store of value that Blockchain technology has shown resilience; it has shown real influence.
Innumerable elaborations on the underlying ideas have been spun off.  
Some are foolish; some are fraudulent; some are sensible.
However, it is clear that at the time of writing the world is awash with a soup of cryptocurrency-related businesses, startups, research projects, and initiatives.  
And more generally --- since a cryptocurrency is just a special instance of a distributed ledger --- the cryptocurrency soup is just one well-known eddy in a cauldron of blockchain-based distributed ledger ideas.

\subsection{Our question is: why?}

This is not to ask just why blockchain has seen growth and resilience --- though this is part of the question.   
Our question is: why has blockchain-based tech \emph{in particular} seen this interest?

There is no shortage of plausible future tech: biotech, fintech (financial technology), quantum tech, 3-D printing tech, battery tech, space tech, renewable energy tech, machine learning tech \dots and so forth.  
It is unclear that blockchain technology will be more important for human progress, or offer a clearly better return on investment (ROI), than (say) biotech, fintech, or machine learning.  
Indeed these are all experiencing their own booms --- and deservedly so, because they have a pedigree of delivering concrete benefits and profits.

What does blockchain have, that (for instance) machine learning does not?

\subsection{A word on terminology}

We may use the terms \emph{blockchain} and \emph{distributed ledger technology} (DLT) interchangeably, but in fact they mean slightly different things: 
\begin{itemize}
\item
a \emph{blockchain} is a (usually distributed, usually cryptographically assured) chain of blocks (the technical term is \emph{Merkle tree}), whereas 
\item
a \emph{distributed ledger} is a database that exists on (i.e. is distributed over) multiple locations (but not necessarily secured on an actual blockchain).
\end{itemize}  
So technically, the git version management system is a blockchain, and a RAID 1 hard drive array is a distributed ledger.
In practice, the terms blockchain and DLT are used to refer to a cryptographically secured block-structured distributed ledger, usually with some element of peer-to-peer communication and consensus.
If we write `blockchain' or `DLT', we mean one of them --- along with a general hint that DLT is intended in a slightly more general sense than blockchain.
A \emph{cryptocurrency} is then a currency whose ownership ledger is distributed on a cryptographically secured blockchain.

\section{A carnival of incompetence}

Blockchain has a history of error, fraud, and incompetence so outrageous\footnote{Thanks to~\cite[page~36]{gerard:attffb} for collecting many of these examples.} that even for the cynical reader it is worth recalling a selection of examples: 
\begin{itemize}
\item
\emph{2011}\quad Bitomat kept its wallet on an ephemeral Amazon EC2 server.  
Being `ephemeral' meant that when the server restarted, it would restart with empty discs.
They restarted the server.  17,000 BTC got wiped.
\item
\emph{2012}\quad Bitcoinia was a 16-year old's first serious PHP project.  It quickly got hacked, and because the admin had reused their password on other secure sites, this led to a wave of further hacks.
18,547 BTC lost.
\item
\emph{2011-2014}\quad 850,000 BTC were siphoned out of the Mt Gox exchange. 
\item
\emph{2015} \quad AllCrypt was run on a MySQL server that also ran WordPress. 
This is the security equivalent of building a bomb shelter with patio doors.  It did not end well.
\item
\emph{2016}\quad Loanbase did the same thing.
\item 
\emph{2017} \quad Zerocoin had a basic error in its code; {\tt ==} (is equal to) instead of {\tt =} (make equal to).  
Coins could be spent twice, and were.
\item 
\emph{2019} \quad QuadrigaCX founder Gerald Cotten died (or possibly absconded), leaving his laptop with 190 million USD in crypto assets locked under a password that only he knew.
\end{itemize}

\section{Q. What is DLT good for? A. Not much, and yet\dots}
\label{sect.good.for}

The list above, tragicomic as it is, cautions against dismissing the interest in blockchain tech as a mere mania.
Yes it is a mania, and full of fools, but if this were only about manic foolishness then presumably --- after ten years' nonsense like the above --- the field would fall into disrepute and the fools would move on to the next thing.\footnote{The honest and competent operator will find that it is technically very difficult to deliver blockchain safely and usefully to end users (so while carefully trying to do so, risks being shoved aside and crowded out by dishonest, incompetent ones).  There are deep reasons for this, which we discuss in Subsections~\ref{subsect.the.tech} and~\ref{subsect.prediction}.}

Instead, interest in DLT has only increased.
After each disaster the investors piled back in, with renewed determination.
Blockchain projects have multiplied and the headline value of Bitcoin has failed to go down the drain.

Yet we struggle to find use cases for DLT on a scale to justify such optimism.
This is not to say that \emph{potential} use cases are inconceivable or elusive: on the contrary, plausible \emph{potential} use cases abound.
What is difficult, even today, is to find \emph{proven} use cases.

In fact there are only two cardinal successes so far, and both are cryptocurrencies: Bitcoin, and Ethereum.\footnote{Apologies to the cryptocurrencies we left out.  At time of writing, Ether's market capitalisation is a tenth that of Bitcoin and it is the largest of the so-called \emph{altcoins} (crypto other than Bitcoin; Ripple is next, with a twentieth).  We mention Ethereum in particular because of its widespread cultural influence, due to its adoption and its emphasis on smart contracts, which has sparked many projects and elaborations \dots}
In terms of demonstrated usefulness and functionality, Bitcoin and Ethereum are of interest only as stores of value.\footnote{\dots but smart contracts, while interesting (especially to a logician/programmer, such as the second author) are not yet a \emph{proven} functionality.  They deserve their own essay which we leave for future work.} 

This is a circular and self-referential answer to the question of why DLT has value: Bitcoin and Ether have value because everyone agrees they have value.
While perhaps satisfactory for an investor, it does not address the fundamental academic question of \emph{why}.

At this point we should perhaps mention that we make no claim that DLT cannot create value, have applications, and demonstrate fundamental ROI beyond Bitcoin and Ethereum as a store of value.

But the intensity and persistence of the mania surrounding this technology, are striking:
a technology with no use cases aside from itself, and with a ten-year history of abject failure, corruption, and people getting fleeced, refuses to die and instead stimulates continued interest and investment.
Investors have not (all) moved on to the next fad.
Researchers (including the authors) remain active in the sector. 
Businesses display patience and continued interest.
 
Why?

\section{The economic ideas of Karl Marx}
\label{economic.marx}

To try and arrive at a fresh understanding of what is happening, we will consider some economic and political theory, and a little history. 

Karl Marx was an economist, political theorist, and socialist revolutionary.\footnote{US lay readers please note: in Europe, `socialist' is not an insult and may even be a compliment.}  
The first two aspects of his work interest us most here.
 
In Marx's seminal essay \emph{Wage Labour and Capital} (\emph{Lohnarbeit und Kapital}) of 1849~\cite{marx:lohak}, Marx defined a \emph{factor of production} as a unit in an economy~\cite{wikipedia:meaop}.
He identified Labour and Capital as two major factors of production:
\begin{itemize}
\item
\emph{Labour} is work: digging ditches, childcare, writing articles, teaching, consultancy, and so forth.
\item
\emph{Capital} is assets: buildings, tools, factories, dollars, bitcoin, rights (such as copyright or land rights), and so forth. 
\end{itemize}
Marx observed that Labour and Capital are fundamentally in opposition:
\begin{itemize}
\item
Labour wants labour to be expensive and assets to be cheap.
\item
Capital wants labour to be cheap and assets to be expensive.
\end{itemize}
This is a simple but not trivial observation: that there is a chain to creating value, and each party in the value chain has an incentive to maximise their benefit. 

For society to function, these imperatives must find an equilibrium: 
\begin{itemize}
\item
Labour --- meaning people who work --- must be paid enough to invest in themselves and their families.

If this were not to happen then society would implode, because nobody would have money to buy capital items, pay for services, and nurture the next generation of educated and productive citizens.
\item
Capital --- meaning people who own stuff --- must be paid enough that owning and maintaining stuff is worthwhile.

If this were not to happen then society would implode, because nobody would have an incentive to take care of things, nor invest in the future to maintain and develop the tools, tangible and intangible, which humans need. 
\end{itemize}
Marx's basic argument was that this equilibrium is unstable and unsustainable: 
\begin{itemize}
\item
Competition would lead to increases in worker efficiency.  
\item
For an individual worker, greater efficiency is generally good --- an internet connection, for instance, helps these authors to efficiently research and teach. 
This is a good thing, which e.g. the authors' students value, and the authors' university employers can monetise.
\item 
However, for the system \emph{as a functioning whole} the effects need not necessarily be so positive: if we can obtain the same productive output from fewer workers, then the value of work overall may drop, unless more work can be invented. 
\end{itemize}
In Marx's day, `work' meant digging ditches (to put it crudely).\footnote{It would be unfair to fault Marx for failing to anticipate the digital economy.  More on this in Section~\ref{sect.digital.economy}.}
But,  the world only needs so many ditches.
Therefore, if e.g. thanks to a mechanical digger the same number of ditches can be dug by 10\% of the workforce, and suppose also that with new teaching technology universities can teach the same number of students with 10\% of the professors --- then 90\% of this workforce becomes redundant.
This is not just in terms of a particular individual no longer having a specific employment, but in terms of the systemic economic relevance of including working people in the system. 

They are not needed; and things that are not economically needed, have no economic value.
Even if we put on our callous capitalist hats and set aside the inhumanity and social damage of the unemployment experience, this also destroys the value of capital itself: because, if the people are not needed, then neither are the capital assets to maintain them as humans.

Thus if the value of labour collapses, then so does the value of capital, and Marx's analysis of the system as a whole, viewed as an economic machine inhabited by humans,\footnote{\dots but not serving them; we have our \emph{callous} capitalist hats on here, for the sake of this argument \dots} is that it will increase in efficiency, shed labour, shed capital, implode, and collapse.

In short: Marx's observation was that the capitalist system, in and of itself and even taken purely in terms of its own internal logic, is unstable and inherently self-contradictory.

The historical record is unclear whether he was right.
Capitalism has not collapsed yet, but that does not mean it won't.
We certainly have no grounds to be smug.\footnote{At this point this article could suggest that a more socially conscious and humane economic system might also be a more stable and sustainable one.
This suggestion is reasonable and worth exploring (and has been explored by other authors \cite{mazzucato:valemt}) --- but not in this essay. We need to follow a different thread.}

We will argue that in a sense, the remarkable recent persistence of interest in DLT reflects an awareness of, and is a response to, the problems described above; but we are not yet ready to describe how.
First, we must survey some recent economic history. 

\section{Labour vs. Capital: a tug of war}

A simple test to see where the balance of power lies between Labour and Capital is to look at inflation:
\begin{itemize}
\item
If inflation is high, Labour is powerful and can command ever greater wages.  
\item
If inflation is low, Capital is powerful and can hold wages down. 
\end{itemize}
From a perspective of Marxian economics, inflation is quite a reliable indicator of the balance of power between Labour and Capital.
If this balance of power shifts too far in one direction or another, central banks can try to manipulate the resulting inflation, or lack of it, by controlling the money supply.  
However, since the financial crisis of 2008, attempts to create inflation have been ineffective --- which suggests that the traditional (monetarist) view of inflation may not be so applicable to a world with globalised finance and supply chains~\cite{elerian:onlgt}.\footnote{So that's \emph{two} things a capitalist's money can't buy: love, and inflation.} 
 
With this test in mind, we will sketch the context of the current interest in blockchain tech, by looking at the broad sweep of inflation over the past eighty years since the end of the Second World War.\footnote{This analysis is US/UK-centric, which is a simplification but a reasonable one, because the dollar is the world's currency and because thus far the main drivers of blockchain technology have also been US/UK-centric.}
We will split this into three eras: 
\begin{itemize}
\item
Era 1, 1945--1979:\ from the end of the Second World War to the elections of Margaret Thatcher (1979-1990) and Ronald Regan (1981-1989).
\item
Era 2, 1979--2008:\  from the Thatcher/Regan era and the corollary administrations which followed, through to the 2008 financial crisis.
\item
Era 3, 2008--today:\  the reality our teenage University students enter as working adults. 
\end{itemize}
In 1945 both Labour and Capital were significantly depleted after the disasters of the first half of the 20th century.
Social solidarity in their wake was high, and it was agreed that both Labour and Capital required investment.

Capital was scarce, but Labour was in even greater demand because: 
\begin{itemize}
\item
labour is required to produce capital, but capital is not required to produce labour,\footnote{Raising children is necessarily labour-intensive, but not \emph{necessarily} capital-intensive.} and 
\item
with depleted capital, labour was relatively unproductive and therefore more of it was required.
\end{itemize}
The 1950s and 1960s were good times to be at work, and because there was plenty of demand for labour to go around, workers could build capital and increase efficiency and see this reflected in rising paycheques.

And, it is not just about wages: society valued Labour and was willing to speculatively invest in skills and education.
For instance, the Servicemen's Readjustment Act of 1944 (the G.I. Bill) was described as a reward for war service --- which it was --- but its deeper meaning was an investment in labour skills and education (for example, free university education). 
Another influential example of post-war investment in Labour has been the UK's Open University (1969), which has inspired open universities in many other countries, e.g. in Israel (1974)~\cite{list-of-open-universities}.

Jumping ahead just for this paragraph from the US in 1944 to the UK in 1998 (in the middle of our Era~2 above): university education, which until then had been free, became not-free as tuition fees were introduced.\footnote{\dots by a UK Labour government, we might add.}
These were modest at first but the precedent had been set and fees rose to a maximum of 9,000 pounds a year in 2010 (this was after the financial crisis, at the start of our Era~3).

In our analysis this social signal should alert us that something important has happened, because tuition fees are --- in fact --- not tuition fees at all, but instead they are a targeted tax on Labour; workers are welcome to join the workforce and add productively to the economy, and all society asks of them in return is to pay a modest upfront tax of 36,000 pounds (assuming a four year course) against future earnings.  What could be regressive about that? 
The contrast with the Era~1 attitude, could hardly be more stark.\footnote{This is in England.  Scotland has no tuition fees.}

In the 1970s something changed: wages stopped rising.
This was a decade of strikes; Marx would say that productivity gains had finally started to eat into the value of Labour.
Fewer workers were required, but they continued demanding the wage increases of previous decades, but now, just as Marx predicted, Labour lacked its old leverage.
Social unrest followed.

Something had to give, and between 1979 and 1981 the balance of power swung to Capital --- which was terrified of inflation, this being one thing that can really eat into the value of a capital asset.

An ideology was required to justify the falling value of work and workers, and this was promoted as \emph{individualism}.
Exaggerated symbols of independence developed, such as Madonna's \emph{Material Girl} and Aronold Schwartzenegger's body.
These myths were of people who didn't need anybody's help and were strong and rich and successful because of it; and they were held up as icons of a new capitalism.  
But the actual effect was to break the power of Labour, since
if everybody is in it for themselves, then they are not in it for one another and they can therefore collectively be exploited more, paid less, and/or fired --- and, as per the ideology, it's their own fault for not working hard enough, not being entrepreneurial enough, and not wanting it enough.\footnote{To be clear, Thatcher and Regan were trying to solve real problems.  
Labour \emph{was} worth less; the strikes could \emph{not} continue; and something \emph{did} need to be done.  

However, the human cost of the adjustments was lasting and non-negligible, and this cost fell mostly on Labour instead of Capital.  
The way the ideology blamed its victims for their victimhood, was as elegant as it was cruel, and the repercussions and costs of this injury and insult resonate to this day --- and not only for Labour:
once asset ownership was more important for income than the labour contributed, this devalued the caring professions like childminders, doctors, and nurses.
Now the coronavirus pandemic (which flared up during the preparation of this article) has thrown a particularly graphic light on how professions that are economically irrelevant as measured by GDP, can become a matter of life and death, and how even if we only care about GDP, failing to account for the value network in which that one economic measure is embedded can become self-defeating.

In retrospect one could certainly try to devise exits from the 1970s situation which in the long term might have been more equitable and economically beneficial. 
 
This concern is relevant today because we are now seeing elements of the strategy replayed; e.g. with the so-called \emph{gig economy}, and individualistic slogans such as \emph{be the CEO of you}.  
These are not necessarily all bad --- a gig may be a welcome earner for e.g. an individual with some energy and free time --- but we have seen in our lifetime what the systemic consequences can be of promoting as a cool lifestyle choice a business model that is great for Capital and terrible for Labour, and can therefore not claim ignorance this time around.} 
The Thatcher-era right-to-buy of council properties~\cite{murie:rigtb} was part of this too: making it possible for workers to take ownership of their homes turned them into holders of capital, and thus gave them a stake in the new system. 

Over the following thirty years to 2008, the value of capital shot up.
In retrospect we see that this postponed the problem but it did not solve it, and the financial crash of 2008 was of a piece with the strikes of the 1970s:
\begin{itemize}
\item
In the 1970s the return on investment in Labour either flattened out or became self-defeating.
In a rush to maintain returns the system became unstable.
With Labour weakened and discredited, political power shifted to Capital.
\item
In the 2000s a similar process took place with Capital; the return on investment in capital flattened out or became self-defeating, and in a rush to maintain returns, the system became unstable --- the precise mechanism was that Capital borrowed money and leveraged it, but the deeper driver here was that Capital had run out of ways to earn real returns, and so it invented imaginary ones. 
\end{itemize}
We now see the problem: where can the pendulum swing next?

\section{Nostalgia vs. the future}

When society is in difficulty, people can either be realistic and analytic about their difficulties \dots or they can look to vivid fantasies of escape.
These fantasies are quite revealing:
\begin{itemize}
\item
\emph{Go back to the 50s and 60s!}\ 
White men were real Men, the world was ruled by Christians, and women and children and black people knew their place. 

Much of current US politics has retreated to this space.
\item
\emph{Go back to the 70s!}\ 
Workers united to face down evil capitalists.  Yes, it didn't come out so well for the workers then, but this time will be different.

This current has not met with electoral success in either the US or the UK, though it may remain relevant as a pigeonhole into which to stereotype proponents of a more socially conscious capitalism.
\item
\emph{Go back to the 80s!}\  
This is like going back to the 50s and 60s, but with more tech and funkier suits.

Baby-boomers are particularly prone to this fantasy, because the 80s were good to them, especially if they bought property.
\item
\emph{Anarchy and authoritarianism} are all jumbled up in a single item of political opportunism.

This would be amusing, were it not for the election of so many outrageously unfit politicians in 2016-2018, along with an enabling retinue of crooks, fanatics, and thieves, all anxious to carve up what is left of the pie before (they fear) the music will finally stop. 
\item
\emph{Zoom to the future, usually with new tech --- like in Star Trek!}\ 
But perhaps without the aliens, or the warp drive, but the point is: tech will solve our problems.\footnote{Marx would have countered that tech \emph{is} the problem.  We return to this in Section~\ref{subsect.the.tech}.}

The dreams of blockchain lie here.
Viewed in context as an escapist fantasy, they are not clearly more dangerous than the alternatives --- and at least as a fantasy it is forward- rather than backward-looking, which matters.
\end{itemize}
It is now important to appreciate that blockchain is not just any old techno-utopian escapist fantasy.
It is also a concrete technology, with specific technical qualities, which make certain promises, not all of which are unreasonable.

Consider what Labour and Capital most fear:
\begin{itemize}
\item
Labour worries that resources will continue to be co-opted by Capital.
This is valid since it has been a pattern for the past forty years --- that is, wealth inequality has trended up \cite{roser:inci,how-has-inequality-changed}.
Most recently the US leaned shamelessly into this with the Tax Cuts and Jobs Act (TCJA) of 2017, in which wealthy Americans essentially agreed to cut taxes on the wealthy and thus allocate more of the pie to themselves. 
\item
Capital worries that Labour, fed up with stagnant wages and aggravated by regressive taxation and lagging and biased social investment, may decide the game is hopelessly rigged and rise up and confiscate assets --- either through progressive taxes and political reform; or through theft and destruction.
Or, perhaps a corrupt government will take advantage of social instability and loss of legitimacy to concentrate wealth in a new elite, as has happened in history before.
These fears too are reasonable.
\item
Both Labour and Capital, again with good reason, fear a currency collapse whose real costs they may be unable to avoid, or inflict upon somebody else.
\end{itemize}
Here, Bitcoin makes some relevant promises: in particular, it promises anonymity and independence from government.
Thus:
\begin{itemize}
\item
Labour projects onto this an ideal of a libertarian currency that is accessible to all, cannot be coopted, and (implicitly) promises financial independence from a corrupted and discredited elite.
\item
Capital projects onto this an ideal of a stable and secure asset that cannot be inflated away or confiscated by a vengeful mob come to grab their share.
\item
Both Labour and Capital like the idea that Bitcoin cannot be inflated away.
One Bitcoin will still be one Bitcoin tomorrow \dots whereas a dollar may be worth only fifty cents. 
\end{itemize}

\section{Blockchain's promise}

In fact, Bitcoin is not anonymous~\cite{bitcoin-not-anonymous},\footnote{It is pseudonymous, which sounds like `anonymous' but means something quite different.}
patterns of Bitcoin ownership resemble those of other capital assets~\cite{state-of-bitcoin}, and Bitcoin is volatile, liable to theft, and difficult to transact in.

\emph{However} fiat currencies share many of the same features: they can also be not anonymous, unevenly distributed, volatile, and liable to theft.
Libertarians particularly emphasise here institutional theft by inflation --- it is unclear how rampant theft of cryptocurrencies by hackers is any better, but at least it's not by a \emph{government}.  
As Satoshi wrote:
\begin{quote}
\emph{The root problem with conventional currency is all the trust that’s required to
make it work. The central bank must be trusted not to debase the currency, but
the history of fiat currencies is full of breaches of that trust.”}
\end{quote}
So in relative terms Bitcoin may not seem that bad, and in any case the utopian \emph{promise} is there, even if the reality has not caught up just yet.

And here is why cryptocurrencies have captured the collective imagination.

It is not just that we live in uncertain times. 
There seems a sense in which we may be at the end of an economic arc, which started in 1945, inflected in 1979, and has reached some endpoint since 2008, where increases in productivity of Labour and Capital have stalled or become counter-productive \emph{at the same time}.

There seems nowhere to turn, the system is wobbling, and the key component of \emph{trust} in that system, is ebbing away.

If so, then this is an arc which Marx predicted. 
The endpoint of his prediction was a social collapse which may yet happen, and (so the fear) we may be trapped on this trajectory by an economic logic which we struggle to escape.

Fantasies of escape to a golden age are obvious nonsense, and even their proponents, wilfully blind as they may be, are aware of this at some level, which accounts for the stridency of much of current politics. 
Next to this, cryptocurrencies promise a way out which is not obviously any more crazy than anything else, and \emph{this} is why Bitcoin has bounced back from one disaster after another, and why research and investment continue to flow to blockchain tech, trying to make it work.

Yes, many of the other techs from Section~\ref{sect.bitcoin.and.dlt} have proven capable of generating ROI within the current system --- but 
as Marx argued, beyond a certain point just increasing the productivity of individual components in an economy does not necessarily make the economic system as a whole stable or more productive, so just delivering ROI does not in itself address the deeper worry.

What no tech can currently promise is an escape from this contradiction, which Marx observed embedded in the system itself --- no tech, that is, except for blockchain.
That promise, however tenuous, is specific to our time. 

The problem is, blockchain technology does not actually deliver.
It is, for now, mostly vapourware (cf. the discussion in Section~\ref{sect.good.for}).

So in summary so far:
we have argued that blockchain promises, but does not yet deliver, a modern technological fix to a specific social and economic problem which was predicted by Karl Marx back in 1849 and which, arguably, the world economy has been stuck in for the decade since 2008.

We will argue next that since Marx's time there have been two developments which he could not have forseen and which may have modified the rules of the game:
\begin{itemize}
\item
The rise of the digital economy, and specifically of intangible assets that can be infinitely copied at near-zero cost, and which must therefore be protected by copyright.
\item
The rise of Data as a new factor of production. 
\end{itemize} 
We will discuss the technology first.

\section{Blockchain as a technology in the digital economy}
\label{sect.digital.economy}

\subsection{The tech}
\label{subsect.the.tech}

For Marx, technology was as much a problem as a solution.
As discussed, it increases worker efficiency to the point that having workers becomes systemically unnecessary, so that capital becomes unnecessary, so that the economic system implodes.

On the face of it the digital economy makes this worse than Marx could have imagined.
Moving electrons around is extremely low cost; it is no harder to make a million copies of a program than one.
A small number of good programmers can move many electrons, and one reason for the concentrations of wealth which we have seen\footnote{Microsoft, Apple, Amazon, and Alphabet were worth roughly 1 trillion USD each in January 2020, and Alibaba, Facebook, and Tencent about half a trillion USD~\cite{market-cap-list}.  That really is a lot of money.}
is surely just down to social policy failing to distribute the exceptional productivity of computer programs.

A concrete example gives a feel for the numbers that can be at play: in 2017 Facebook bought WhatsApp for 19 billion USD.
At the time, WhatsApp had 55 employees.
Some of this money was set against projected future earnings, but the fact remains that on a per-capita basis, the market judged that each WhatsApp employee had created 350 million USD in current or future value.

For this reason Capital loves the digital economy.  
Once you have copyright control of a program or film, or a patent, you have an economic perpetual motion machine and need do little more than sit back and rake in the profits.\footnote{%
\emph{Copyright trolls} practice the purest form of this business model, incurring just the fixed costs to acquire the patent and then enjoying very low marginal costs.  This impressively combines being a malevolent and parasitic modern piracy, with being an elegant distillation of an essential truth of the digital economy.}  
Labour is effectively removed from the equation --- aside from perhaps having to feed and clothe the families of no more than 55 programmers. 

Labour also loves the digital economy.  
It brings comfort and increased productivity --- provided you have the necessary skills, of course.
If not, you may have a problem.\footnote{\dots and so may society.  We still need waiters, nurses, teachers, and childminders.  The problem is figuring how to pay them, which was part of Marx's critique.}

`Technology' however is a broad term, and blockchain tech has some quite unusual features which may not be generally appreciated.

Let us look at the core technology that had to be harnessed in order to start off the revenue streams of some companies that are now worth hundreds of billions of USD:
\begin{itemize}
\item
Facebook is fundamentally based on a labelled graph.
\emph{Graph} here is in the mathematical sense of a connected object with \emph{nodes} (dots) joined by edges (lines); 
a Facebook landing page, with icons of people on a map of the world and connected by lines, sums it up. 

The mathematical notion of graph is a first-year undergraduate topic: it is not hard.
\item
The Google pagerank algorithm is, fundamentally, an advanced undergraduate project or perhaps early-stage PhD.
\item
Amazon is `just' a database, and MS-Dos, Windows, and MS Word are fairly simple programs. 
\end{itemize}
This is not to say that Facebook, Google, Amazon, and Microsoft are based on simple systems; on the contrary, scaling up from the initial product to the worldwide firms we know today is technically very complex.

Nor is this to say that creating such companies was simple and easy. 
Far from it; there were more failures than successes.

Our point is that, if we assume the course is set, the winning formula found, and that whatever wisdom or luck that was required has been added to the mix --- in other words, if we assume we have won this particular lottery --- then the fundamental \emph{conceptual} requirements of getting the tech started, were not great.

Contrast this with for example ARM, Intel, and Airbus.
The tech required to fabricate chips or aircraft is rocket-scientist level from the start; it requires specialists with years of training, and specialist managers able to manage them and if things go wrong then people get very excited and have government inquiries.\footnote{Yes, the Wright brothers worked out of a bicycle shop, but that was over a hundred years ago and they were aiming for a prototype plane, not a socially transformative global cryptocurrency.  The two things are different.}  

To put it another way: you can't start a chip fabrication business in your garage, but you can start a trillion dollar software company --- or at least, you could ten or fifteen years ago (nowadays the incumbents would buy you up or crowd you out, but that is a different debate).

Blockchain in its infancy is more like Airbus than Amazon.
To start a revenue stream based on DLT requires expertise, research, and special management which simply has not yet been assembled in the sector.
Consider that a company wishing to generate a revenue stream from a cryptocurrency product must commit to solving mathematical problems that are:
\begin{itemize}
\item
fundamentally hard,
\item
in a safety-critical context,
\item
for consumer use.
\end{itemize}
It is ambitious, and may be impossible, to satisfy all three of these criteria simultaneously in an \emph{initial product}.

Part of the difficulty is that this requires specialised programmers and mathematicians to invent and implement a new body of mathematics, but also, it requires processes and qualified managers who do not exist yet because the field has not matured; potential regulatory changes to provide a legal framework into which users can escape if and when things go wrong (because they always do); and a programme of public education to educate a population who (as a general body) may still be storing their PaS5w0rds on post-it notes stuck to the monitor.

\subsection{A prediction}
\label{subsect.prediction}

Based on the analysis above, we would like to offer a prediction, just for the sake of argument.

The killer apps and real impact of blockchain will not appear first as front-end consumer applications.\footnote{This is not to say that consumer applications will not exist; just that they will be made for the sake of it and not actually be that important, except perhaps for public education and diffusion of the ideas.}
They will appear in backend applications, developed by large institutions (banks, or logistics companies, or large tech companies) for internal use or for use with their institutional peers.

These will be organisations with deep pockets and long experience of managing extreme complexity, who can afford to engage (and if necessary to create) a small army of highly specialised programmers, lawyers, computer scientists, and mathematicians to develop a product that requires a PhD just to switch it on safely.

In other words: in spite of the success of Bitcoin we predict that blockchain tech will behave economically more like rocket science, and less like cash, and will democratise only later, if at all. 

\subsection{A few words on data}

On the face of it, Data is just an asset.
By this view, owning data is like owning any other capital asset. 

However, Data has the potential to become a factor of production in its own right; an actor on a level with or even ruling over Labour and Capital.

What is the difference?  In practice, we propose that the difference between an asset class and a factor of production is that the former only demands maintenance, whereas the latter tends to actively pull itself together into a self-serving entity:
\begin{itemize}
\item
Labour is people.
People flow together because humans are social.
\item
Capital is ownership of things.
Things demand care to remain operational, increase in value, and not be stolen.
This gives holders of capital as a group an incentive to flow together to cooperate and protect their wealth.
\item
Data in its modern forms, looks like an asset for input e.g. to machine learning and AI algorithms, but it is not just that: Data requires curation to harvest and clean it, infrastructure to process it, and it displays potent networking effects; a database that is twice as large is (crudely put) four times as useful.
These properties give data a form; it tends to want to flow together into a single entity, just like Labour and Capital do.
\end{itemize}
It will be a social and regulatory question whether Data becomes a capital asset, or a factor of production on a par with Labour and Capital, and if the latter, whether it will be an equal, a subordinate, or the master of Labour and Capital. 

This is relevant to our Marxian analysis because it is one pertinent way in which the world which Marx was thinking in, differs substantively from the world which we now inhabit.
There may be a new actor in the drama, and how they could influence the story is not yet understood. 
We leave a fuller analysis to future work.

\section{The way ahead}

If Marx were around today, we suspect he would have liked crypto, at least in principle.
The promise of Bitcoin to allow Labour to mine coin and thus take control of the means of production of what one might call \emph{cryptocapital}, is a relevant\footnote{The actions of the US Federal Reserve in the coronavirus pandemic have reminded us how the means of production of capital in the US (and thus in the world) are operated to serve an oligarchic elite.  It would not be prudent to stake one's life and livelihood on an expectation that Capital will share the pie, because it won't.} and clearly Marxian dream --- though sadly it is \emph{only} a dream.\footnote{Mining Bitcoin has long ceased to be democratic; most is now mined in huge centralised `Bitcoin farms', or by distributed malware.  

For the record: Bitcoin is a proof-of-work system (proof-of-stake is inherently different). 
The industry is aware of this issue and attempts are being made to design cryptocurrencies that do not reward centralisation so much.
(If the reader has tried to buy a graphics card in the past five years, that price spike you may have noticed was due to crypto mining --- and now graphics cards are used for AI.  But we digress.)}

By looking at the world as Marx did in 1849, we have obtained an analysis of why blockchain tech has been so uniquely resilient in 2008-2020, and what it really promises for the future: namely, an escape from Marx's logical contradiction as outlined in Section~\ref{economic.marx}.

We can now turn to a natural corollary question: could Blockchain actually save us from the trap that Marx identified?
Perhaps.

In Blockchain's favour, it does not generally destroy jobs, and it renews incentives to invest in Labour, because designing, implementing, and managing these systems is skilled work, in highly specific ways.

Making blockchain-related technologies work implies significant investment in logic and formal methods;\footnote{This is the second author's professional specialisation.} tens and perhaps hundreds of thousands of highly qualified programming and management jobs; the democratisation of high-level, high-assurance programming languages and the skills to use them; much work in communication and education; entire new areas of mathematical research, regulation, and legal specialisation; and arguably, the rise of a new breed of hyper-technical companies employing programmers with skills that are currently expected only of the highest-level graduates.
In other words, Blockchain means employment, innovation, and investment in skills, and this will increase the value of Labour.

Importantly, Blockchain is at the same time creating capital, albeit of kinds not seen before --- including but not restricted to Bitcoin.
This may relieve some of the tension between Labour and Capital which (by our analysis) has so destabilised the world for the past ten years.\footnote{Thanks to the reader who astutely pointed out that Marx might say here that this can only be temporary.  No doubt, but a breathing space is still space to breath, and all other things being equal, a technology that promises to create both labour and capital is better than one that does not.}

It is worth pausing to compare and contrast the profile of blockchain tech with that of machine learning / AI tech, which at first glance appears similar in that it is also very mathematical and computer-intensive.
It too requires highly qualified labour, though perhaps not as much of it.
However: the underlying technology does not need to be democratised in quite the same way; its incentive is to resist rather than promote the creation of legal frameworks (which increase costs); and its raw input capital is Data (often from surveillance capitalism) which is \emph{harvested}, not \emph{built}.
In short: there is a possible incarnation of blockchain tech, if it is ever created, which will want a community of active and informed citizens using and innovating with its products to create economic output while protected by well-considered legislative frameworks --- whereas machine learning seems to want to harvest data from a herd of humanity which uses its products via controlled portals in \emph{de facto} walled gardens, while operating in a legal void except for those laws required to protect the incumbents' datasets and IP.\footnote{This is not to say that machine learning and AI are bad --- just that it is not the case that all tech is the same, just because it's tech.  Each technology has its own character, and the gist of this article is an analysis of the specifics of Blockchain tech through a lens of Marx's theories.}

So, it may be that working towards even a fantasy of a blockchain-based solution to the stagnation since 2008, may in itself usefully help to escape it --- regardless of whether it actually works for the purposes originally intended.
At a human and social level, that would be good enough.\footnote{A related discussion of blockchain as a \emph{convening technology} (in the music industry) is in~\cite{swartz:contbm} --- meaning a technology which may or may not actually work on a technical level but which serves to ``galvanize goodwill and to imagine a specific shared potential future, together with implications that have value beyond any ultimate success of the technology around which they convene''.}

A remarkable confluence of technologies is necessary for blockchain to work, including: cryptography, computer science, mathematics, law, communications infrastructure, and public education.
To be useful this combination must diffuse and democratise, and it is impossible to predict the effects this will have --- not just on society, but on the technology as well.
As blockchain spreads, it will evolve.

To democratise and diffuse, this technology will require a broad coalition: from a pair of professors writing in Scotland to an Ethiopian coffee farmer; all parties will be trying to solve their own individual problems, and if the tech is to be truly useful it will need to assume different yet compatible forms, working within sensible social and legal structures yet to be devised.
This democratisation and diffusion is likely to change the tech beyond recognition, and what started as a techno-utopian dream may transition to be the backend of a bank, and then go elsewhere and assume forms that we would hardly recognise.\footnote{A portion of this larger point is made in detail in~\cite{swartz:blodit} observing that the technology may in large part evolve from being \emph{radical} to being \emph{incorporative} (fitting into existing technical structures).}

One thing that seems likely --- and of which we hope that Marx would approve --- is that if we can make this happen, then it will require a lot of work, and a lot of infrastructure.
That would be Labour and Capital working together again --- and that, in and of itself, would be cause for hope.


\begin{thebibliography}{10}

\bibitem{bitcoin-not-anonymous}
Is {B}itcoin {A}nonymous?
\newblock
  \weblink{https://web.archive.org/web/20200628212541/https://bitcoinmagazine.com/what-is-bitcoin/is-bitcoin-anonymous}

\bibitem{swartz:contbm}
Nancy Baym, Lana Swartz, and Andrea Alarcon.
\newblock Convening technologies: Blockchain and the music industry.
\newblock {\em International Journal of Communication}, 13:402--421, 2019.

\bibitem{elerian:onlgt}
Mohamed el~Erian.
\newblock {\em The Only Game in Town: Central Banks, Instability, and Avoiding
  the Next Collapse}.
\newblock Penguin Random House USA, 2016.
\newblock ISBN 978-0812997620.

\bibitem{gerard:attffb}
David Gerard.
\newblock {\em Attack of the 50 Foot Blockchain: Bitcoin, Blockchain, Ethereum
  \& Smart Contracts}.
\newblock CreateSpace Independent Publishing, 2017.
\newblock ISBN 978-1-974-00006-7.

\bibitem{haber:howtsd}
Stuart Haber and W.~Scott Stornetta.
\newblock How to time-stamp a digital document.
\newblock {\em Journal of Cryptology}, 3:99--111, 1991.

\bibitem{marx:lohak}
Karl Marx.
\newblock Lohnarbeit und {K}apital ({W}age {L}abour and {C}apital).
\newblock {\em Neue Rheinische Zeitung}, April 1849.
\newblock Available online in English translation:
  \weblink{https://web.archive.org/web/20200709033651/https://www.marxists.org/archive/marx/works/1847/wage-labour/index.htm}

\bibitem{mazzucato:valemt}
Mariana Mazzucata.
\newblock {\em The Value of Everything: Making and Taking in the Global
  Economy}.
\newblock Public Affairs, 2018.
\newblock ISBN 978-0-241-34779-9.

\bibitem{murie:rigtb}
Alan Murie.
\newblock The right to buy: History and prospect.
\newblock
  \weblink{https://web.archive.org/web/20200807135159/http://www.historyandpolicy.org/policy-papers/papers/the-right-to-buy-history-and-prospect}, 
  11 November 2015.

\bibitem{state-of-bitcoin}
Lucas Outumuro.
\newblock A {C}omprehensive {A}nalysis of {B}itcoin's {C}urrent {S}tate.
\newblock
  \weblink{https://web.archive.org/web/20200707002316/https://medium.com/intotheblock/a-comprehensive-analysis-of-bitcoins-current-state-b41dc2a4dc44},
  July 2020.

\bibitem{roser:inci}
Max Roser and Esteban Ortiz-Ospina.
\newblock Income inequality.
\newblock Our World in Data.\
  \weblink{https://web.archive.org/web/20200701052224/https://ourworldindata.org/income-inequality}, 2013.
\newblock Version of October~2016, accessed 9 July 2020.

\bibitem{swartz:blodit}
Lana Swartz.
\newblock Blockchain dreams: Imagining techno-economic alternatives after
  bitcoin.
\newblock In Manuel Castells, editor, {\em Another economy is possible}, pages
  82--105. Polity Press, 2017.
\newblock ISBN 978-1509517206.

\bibitem{how-has-inequality-changed}
The~Equality Trust.
\newblock How has inequality changed?
\newblock
  \weblink{https://web.archive.org/web/20200624035158/https://www.equalitytrust.org.uk/how-has-inequality-changed}.

\bibitem{list-of-open-universities}
Wikipedia.
\newblock List of open universities.
\newblock \weblink{http://web.archive.org/web/20191002162549/https://en.wikipedia.org/wiki/List_of_open_universities}.

\bibitem{market-cap-list}
Wikipedia.
\newblock List of public corporations by market capitalization.
\newblock
  \weblink{https://web.archive.org/web/20200628042927/https://en.wikipedia.org/wiki/List_of_public_corporations_by_market_capitalization}.

\bibitem{wikipedia:meaop}
Wikipedia.
\newblock Means of production.
\newblock
  \weblink{http://web.archive.org/web/20200701000813/https://en.wikipedia.org/wiki/Means_of_production\#Related_terms}.

\end{thebibliography}

\end{document}